\documentclass[conference]{IEEEtran}
\IEEEoverridecommandlockouts
\usepackage{cite}
\usepackage{amsmath,amssymb,amsfonts}
\usepackage{algorithmic}
\usepackage{graphicx}
\usepackage{textcomp}
\usepackage{xcolor}
\usepackage{booktabs}
\usepackage{listings}
\usepackage{color} %
\usepackage{pdfpages}
\usepackage[hidelinks]{hyperref}
\usepackage{lipsum}  %

\definecolor{codegreen}{rgb}{0,0.6,0}
\definecolor{codegray}{rgb}{0.5,0.5,0.5}
\definecolor{codepurple}{rgb}{0.58,0,0.82}
\definecolor{backcolour}{rgb}{0.95,0.95,0.92}

\lstdefinestyle{mystyle}{
    backgroundcolor=\color{backcolour},   
    commentstyle=\color{codegreen},
    keywordstyle=\color{magenta},
    numberstyle=\tiny\color{codegray},
    stringstyle=\color{codepurple},
    basicstyle=\ttfamily\footnotesize,
    breakatwhitespace=false,         
    breaklines=true,                 
    captionpos=b,                    
    keepspaces=true,                 
    numbers=left,                    
    numbersep=5pt,                  
    showspaces=false,                
    showstringspaces=false,
    showtabs=false,                  
    tabsize=2
}

\def\BibTeX{{\rm B\kern-.05em{\sc i\kern-.025em b}\kern-.08em
    T\kern-.1667em\lower.7ex\hbox{E}\kern-.125emX}}
\begin{document}

\title{Cocobo: Exploring Large Language Models as the Engine for End-User Robot Programming
}

\author{\IEEEauthorblockN{Yate Ge\textsuperscript{1}, Yi Dai\textsuperscript{1}, Run Shan\textsuperscript{1}, Kechun Li\textsuperscript{1}, Yuanda Hu\textsuperscript{1}, Xiaohua Sun\textsuperscript{2}}
\IEEEauthorblockA{\textsuperscript{1}Tongji University, China \\
geyate@tongji.edu.cn, ,2210999@tongji.edu.cn, 2233672@tongji.edu.cn, 2233677@tongji.edu.cn, ydhu@tongji.edu.cn}
\IEEEauthorblockA{\textsuperscript{2}Southern University of Science and Technology, China \\
sunxh@sustech.edu.cn}
}

\maketitle

\IEEEpubidadjcol

\begin{abstract}
End-user development allows everyday users to tailor service robots or applications to their needs. 
One user-friendly approach is natural language programming. However, it encounters challenges such as an expansive user expression space and limited support for debugging and editing, which restrict its application in end-user programming.
The emergence of large language models (LLMs) offers promising avenues for the translation and interpretation between human language instructions and the code executed by robots, but their application in end-user programming systems requires further study. 
We introduce Cocobo, a natural language programming system with interactive diagrams powered by LLMs. Cocobo employs LLMs to understand users' authoring intentions, generate and explain robot programs, and facilitate the conversion between executable code and flowchart representations. Our user study shows that Cocobo has a low learning curve, enabling even users with zero coding experience to customize robot programs successfully.

\end{abstract}


\footnotetext{
    This is the accepted version of a paper accepted for presentation at the IEEE Symposium on Visual Languages and Human-Centric Computing (VL/HCC), 2024.}

\section{Introduction}

\begin{figure*}[htbp]
  \centering{\includegraphics[width=0.9\textwidth]{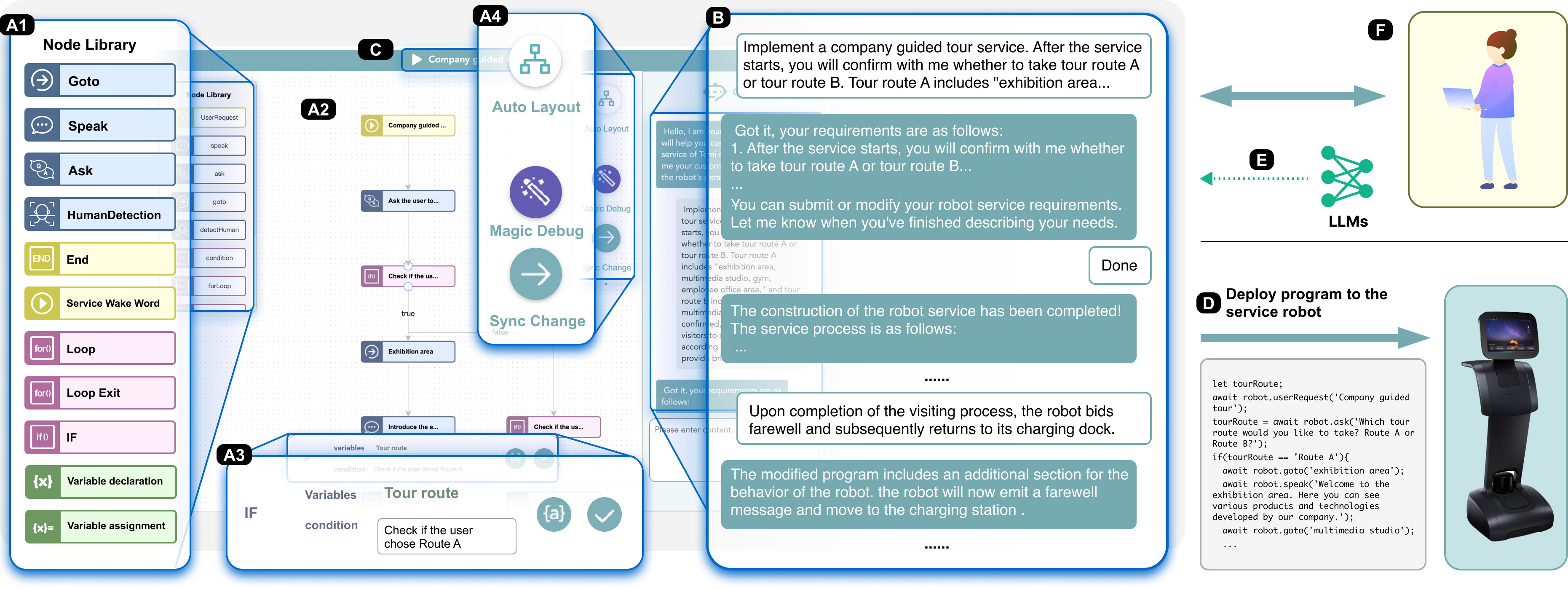}}
\caption{Overview of Cocobo, an LLM-powered end-user programming system for service robot task customization. The system’s UI consists of two views: (A) Flowchart View, which is composed of several components including (A1) node library, (A2) interactive flowchart, (A3) node properties panel, and (A4) buttons group; (B) The Conversational View enables natural language conversational interaction between users and the system.}
\label{fig: overview}
\vspace{-10pt}
\end{figure*}

End-user development (EUD) enables non-technical users to modify applications to meet specific needs, enhancing versatility and user engagement\cite{LiebermanEndUserDevelopmentEmerging2006}. This trend is increasingly relevant in robotics, as buyers now consider software support for defining robot missions a key criterion\cite{Dragulesurveydesignspace2021}. 
The development of mainstream technologies has further stimulated research
interest in this approach\cite{PaternoEnduserdevelopmentpersonalizing2019}, including Internet of Things (IoT)\cite{DesoldaEmpoweringEndUsers2017} and robotics\cite{AjaykumarSurveyEndUserRobot2022}.
Among EUD methods, natural language programming is intuitive but often imprecise, leading to potential misinterpretations between user intentions and program outcomes. Integration with visual programming can mitigate these challenges by helping users better understand and modify their programs\cite{AjaykumarSurveyEndUserRobot2022}.

Recent advances involve pre-trained large language models (LLMs) like GPT-3.5 and GPT-4\cite{BubeckSparksArtificialGeneral2023}, and Llama\cite{ZhangLLaMAAdapterEfficientFinetuning2023}, which excel in code generation and adapting to specific tasks with minimal examples\cite{WeiEmergentAbilitiesLarge2022, DongSurveyIncontextLearning2023}. These models offer the potential for creating more adaptable and user-friendly programming interfaces, especially for personalizing robot tasks.

Research on non-experts’ interactions with Large Language Models (LLMs) in EUD contexts is limited, highlighting the necessity for interactive interfaces that capitalize on LLMs for tasks like the personalized customization of service robots.
This paper investigates how LLMs can assist daily users in programming service robots by leveraging their world knowledge and in-context learning to seamlessly integrate natural language and flow-based programming. This integration facilitates easier program creation, debugging, and modification, thereby lowering the learning curve for users. We developed Cocobo, a proof-of-concept system designed with these capabilities, and conducted an empirical study with 16 participants to evaluate its usability and effectiveness.

\section{Related Work}

\textbf{End-user development for robot task customization.}
EUD empowers non-programmers to modify robot programs, enabling adjustments during runtime rather than only at the design stage\cite{LiebermanEndUserDevelopmentEmerging2006}. EUD methods include visual programming\cite{LeonardiTriggerActionProgrammingPersonalising2019,ErichVisualEnvironmentReactive2017,PotChoregraphegraphicaltool2009,PaxtonEvaluatingMethodsEndUser2018,DattaRoboStudiovisualprogramming2012}, natural language programming\cite{PorfirioSketchingRobotPrograms2023,BuchinaDesignevaluationenduser2016,ErichVisualEnvironmentReactive2017,MongeRoffarelloCorrectionDefiningTriggerAction2023,GorostizaEnduserprogrammingsocial2011}, and programming by demonstration\cite{SaundersTeachMeShow2016}, designed to minimize learning barriers and cognitive load\cite{  CoronadoVisualProgrammingEnvironments2020, GiannopoulouProgrammingItnot2021}. 
However, the required domain-specific knowledge presents challenges, deterring non-experts and increasing the likelihood of errors\cite{OishiEndUserProgrammingRobots2017}. 
Additionally, while prior research often catered to experts, it generally neglected consumer-level users\cite{GiannopoulouProgrammingItnot2021}. The integration of AI for natural language processing and program synthesis offers potential for more intuitive EUD interfaces\cite{FischerAdaptiveAdaptableSystems2023}.

\textbf{Natural language programming with LLMs.}
Natural language programming enhances EUD by allowing code generation via speech or text, supported by AI advancements and large language models\cite{AjaykumarSurveyEndUserRobot2022, WeiEmergentAbilitiesLarge2022}. Despite its potential, using large language models effectively in EUD involves challenges like complex prompt design\cite{SarkarWhatitprogram2022a, SrinivasaRagavanGridBookNaturalLanguage2022, Zamfirescu-PereiraWhyJohnnyCan2023, LiuWhatItWants2023, JiangDiscoveringSyntaxStrategies2022, JiangGraphologueExploringLarge2023, KoSixLearningBarriers2004}. Research indicates that structured prompts could improve robot programming with LLMs\cite{WeiChainofThoughtPromptingElicits2023, KojimaLargeLanguageModels2023, VempralaChatGPTRoboticsDesign2023, KarliAlchemistLLMAidedEndUser2024}. 
Additionally, LLMs’ in-context learning\cite{DongSurveyIncontextLearning2023} capabilities enable them to learn from the context of the task, supporting the generation of multiple representations\cite{Ainsworthfunctionsmultiplerepresentations1999} of programs. This helps to address the uncertainties in natural language programming and supports debugging and editing through structured representations.

\textbf{LLM-based Interactive Systems.}
LLMs enhance interactive systems by enabling conversational interactions and advanced reasoning, significantly streamlining complex task management. Techniques such as prompt chaining break down tasks into simpler, manageable steps\cite{10.1145/3491102.3517582}, exemplified by Metaphorian, which uses sequential LLM prompts to generate metaphors\cite{KimMetaphorianLeveragingLarge2023}. In AI-assisted programming, LLMs improve interaction modalities and optimize information exchange, thereby reducing programmer workload and enhancing efficiency\cite{XuInIDECodeGeneration2022}. These systems develop dialogic interfaces that aid programmers in understanding, debugging, and modifying code more effectively\cite{RossProgrammerAssistantConversational2023a}.

\section{Design and Implementation of Cocobo}

\begin{table}[ht]
    \centering
    \caption{Robot Commands}
    \label{table: robotCommand}
    \begin{tabular}{p{0.35\columnwidth}p{0.55\columnwidth}} %
        \toprule
        \textbf{Robot Command} & \textbf{Description} \\
        \midrule
        userRequest: WakeWord & Activate the service using WakeWord \\
        goto: Place & Move to Place \\
        say: Speech & Say the contents of Speech \\
        ask: Speech & Ask the contents of Speech \\
        humanDetection & Determine whether there is a person in front of the robot. \\
        \bottomrule
    \end{tabular}
\end{table}

We introduced the Cocobo system to assist novice users in creating personalized robot services, detailed in Figure~\ref{fig: overview}.
We selected common robot commands to verify the feasibility of Cocobo, as listed in Table~\ref{table: robotCommand}.
These commands interact directly with the Temi robots’ APIs\cite{robotemisdk2024}.
Further details on Cocobo’s features are discussed in the following sections.

\subsection{The Cocobo User Interface}

Cocobo’s interface includes a conversational UI and a flowchart editor. The conversational UI facilitates the creation or modification of personalized service programs via multi-turn dialogue, while the flowchart UI allows users to check and edit flowcharts. This enables users to understand program behavior in a structured manner and make necessary edits.

\subsection{LLM-powered Interactive Functions}

\begin{figure*}[htbp]
\centering
\includegraphics[page=1,width=0.9\linewidth]{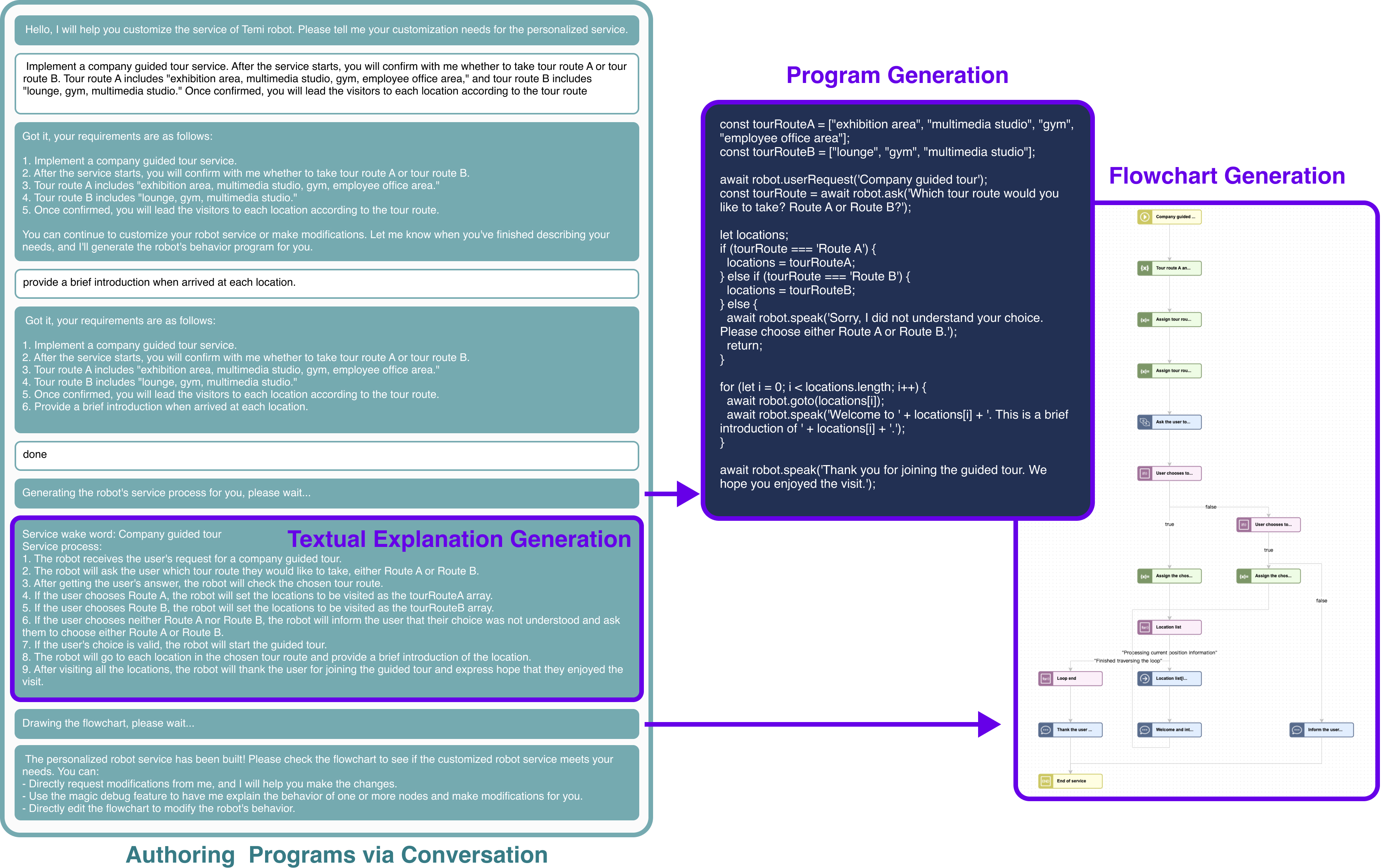}
\caption{Users interact with Cocobo through multiple rounds of dialogues to communicate and customize their requirements. This interaction further facilitates the generation of code, as well as the creation of accompanying textual explanations and flowchart representations.}
\label{fig: authoring}
\end{figure*}

\textbf{Authoring programs via conversation:}
During the authoring phase, users specify their customization requirements through interactions with the dialogue agent.
Following each user input, the system provides a list-based representation of the user’s customized needs derived from its assessment.
Upon user confirmation, the system generates code, flowcharts, and textual explanations(Figure~\ref{fig: authoring}).

\begin{figure}
\centering
\includegraphics[page=1,width=0.9\linewidth]{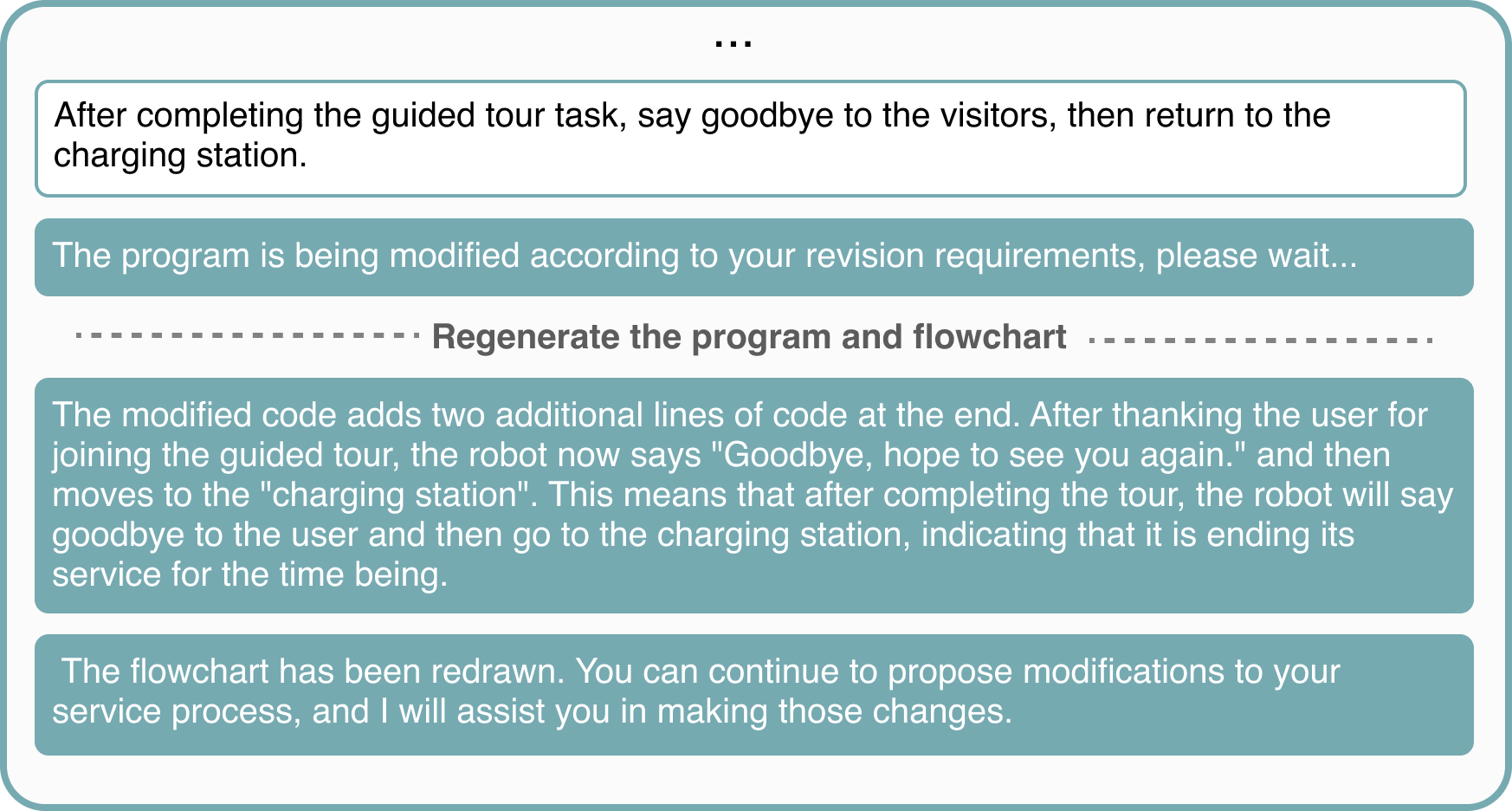}
\caption{Users can directly modify their requirements through text input to alter the code. Cocobo will interpret these modifications and regenerate the flowchart accordingly.}
\label{fig: NLModify}
\vspace{-10pt}
\end{figure}

\textbf{Modifying programs via conversation:}
After the program’s initial creation, users can further specify modification requirements through dialogue. Cocobo regenerates the program and updates the explanations and the flowchart accordingly (Figure~\ref{fig: NLModify}).

\begin{figure}[htb]
\centering{\includegraphics[page=1,width=0.8\linewidth]{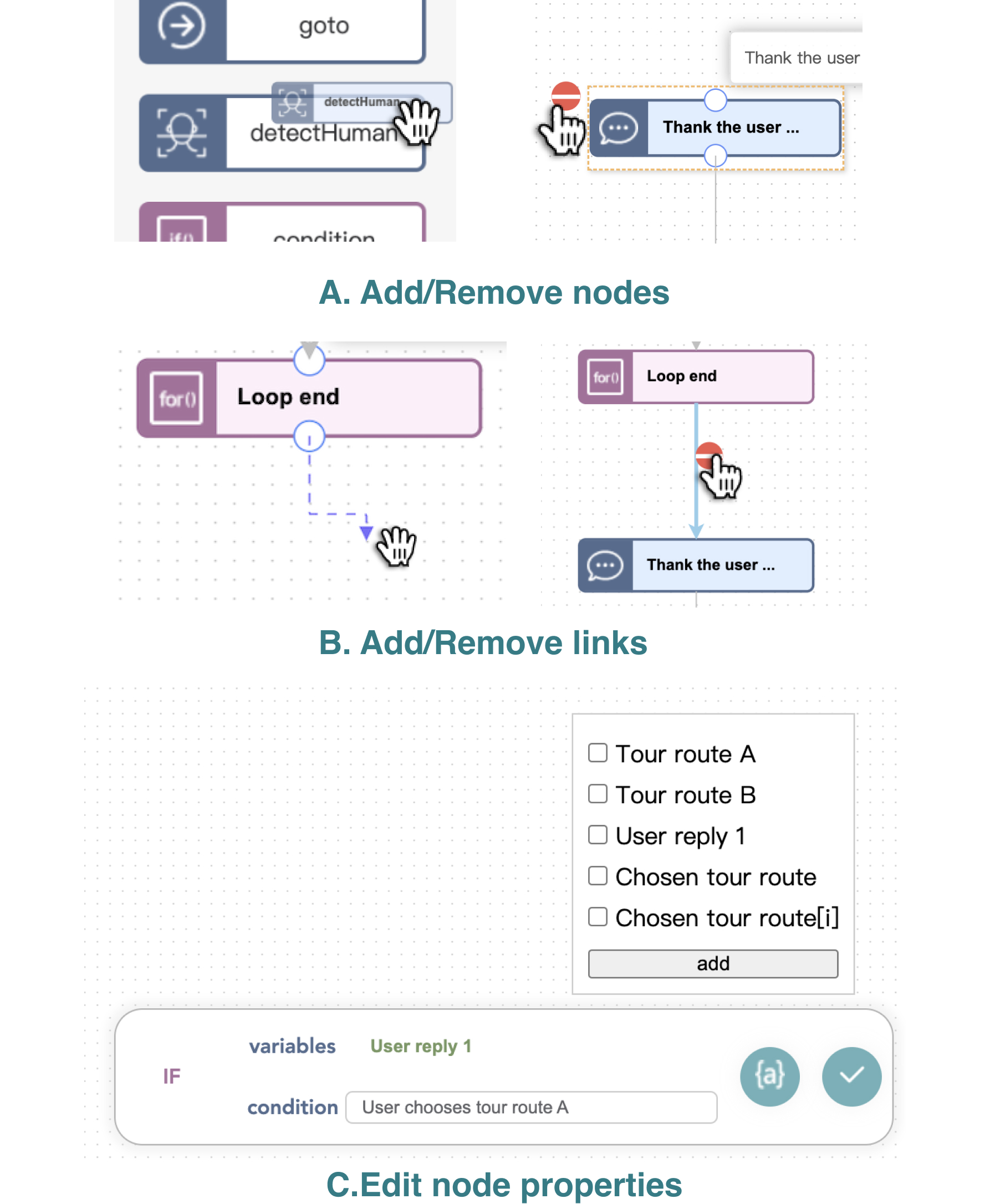}}
\caption{An illustration of the flowchart's editing capabilities allows users to add and remove nodes, as well as modify the connections between them. Within the node properties panel, users can modify the robot behaviors represented by each node using natural language.}
\label{fig: flowchart}

\end{figure}

\textbf{Modifying programs via Flowchart Editor:}
Using the flowchart, users can visually confirm that the program aligns with their intentions and can modify the executable code by adjusting the flowchart. After the user taps the Sync Change button (Figure~\ref{fig: overview}, A4), Cocobo updates the program accordingly. 
Like other node-based visual programming environments, the flowchart editor facilitates adding or removing nodes, adjusting connections between nodes, and modifying node attributes. However, unlike typical visual programming environments, our node attributes feature natural language descriptions of node behaviors rather than direct changes to actual parameters. This approach enables textual modifications to adjust these behaviors (see Figure~\ref{fig: flowchart}).

\begin{figure}[htbp]
\centering
\includegraphics[page=1,width=0.9\linewidth]{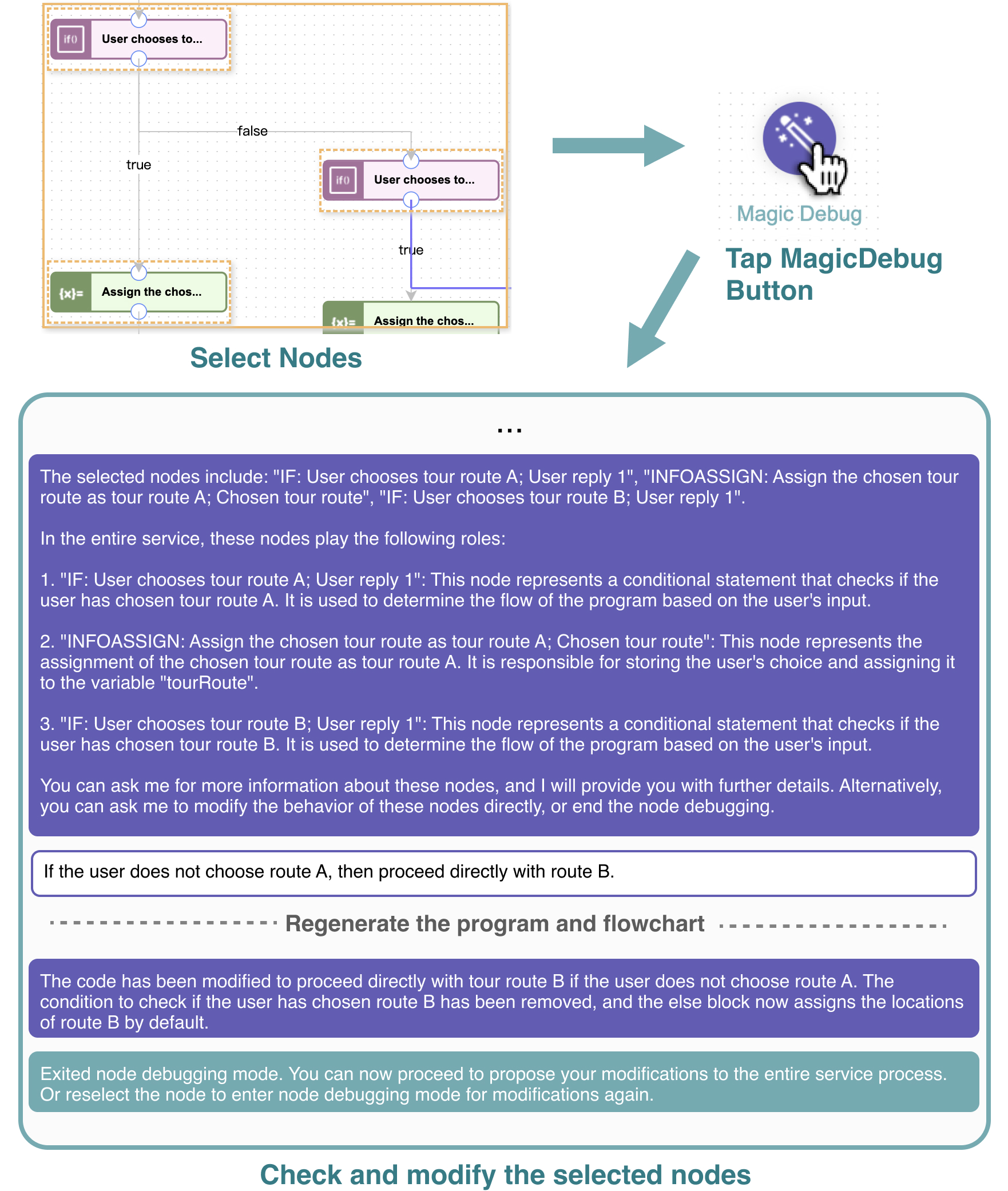}
\caption{An illustration of the usage process for the Magic Debug feature: After selecting one or multiple nodes, users click the Magic Debug button(Figure~\ref{fig: overview}, A4). 
The system then enters node debugging mode, allowing users to inspect and modify the selected nodes through the conversational UI.}
\label{fig: MagicDebug}
\end{figure}

\textbf{Modifying programs via MagicDebug:}
Cocobo features a MagicDebug function that enables users to select nodes and perform targeted debugging and direct modifications via text dialogue.
As shown in Figure~\ref{fig: MagicDebug}, upon user selection of one or multiple nodes and activation of the MagicDebug button, the system provides explanations about the behaviors represented by these nodes via the conversational interface. The dialogue then shifts to MagicDebug mode, enabling further inquiries and modifications to these nodes through text input.

\subsection{LLM Pipeline and Prompt Strategies}

\begin{figure*}[htbp]
  \centering
  \includegraphics[width=\linewidth]{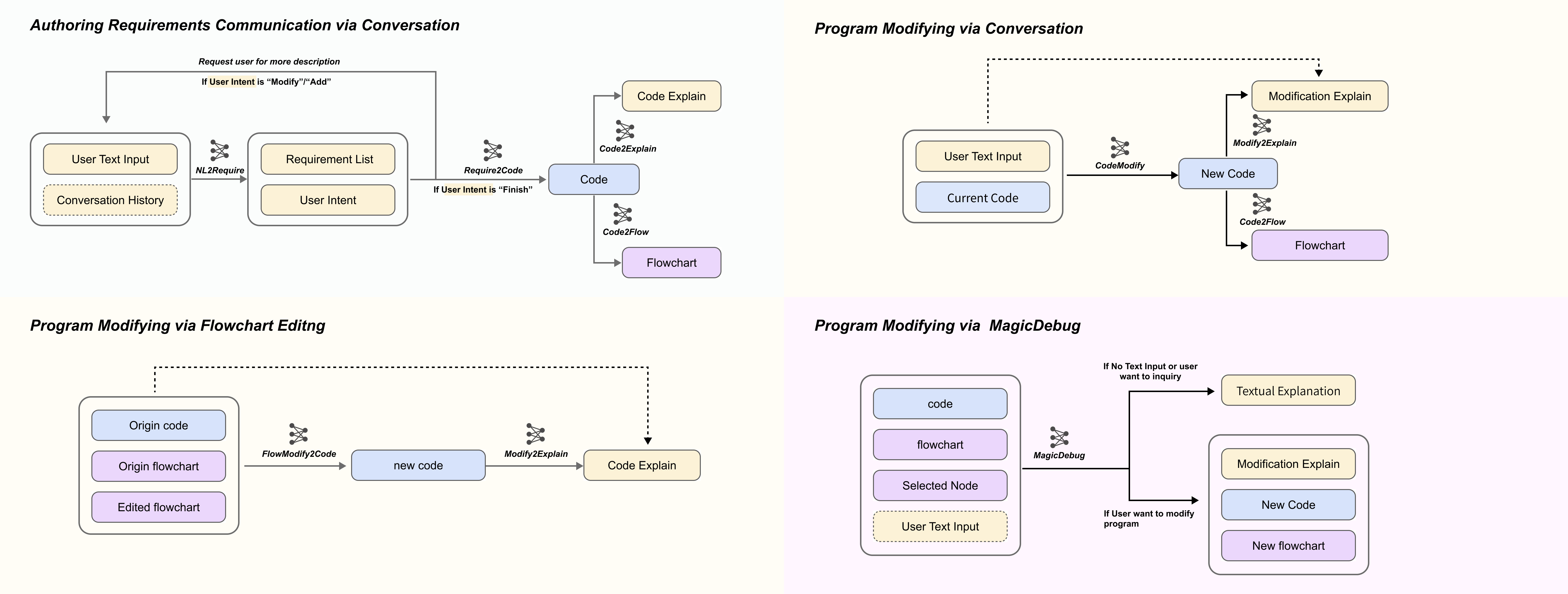}
  \caption{The prompt chaining of each LLM-Powered function in Cocobo}
  \label{fig: systemDesign}
\end{figure*}

As shown in Figure~\ref{fig: systemDesign}, we implemented the core LLM-powered functions of the Cocobo system using the LLM-chain method.
Informed by prior studies on LLM prompting, we adopt a chain-of-thought\cite{WeiChainofThoughtPromptingElicits2023} approach to structure prompt words and decompose tasks in our system. The prompt preamble in our system categorizes prompt words into six key segments: [Role], [Context], [Rules], [Workflow], [Output Format], and [Example]. This structured approach assists in building the LLM module within Cocobo’s LLM pipeline, enabling it to accomplish various tasks.

\textbf{How to enable the system to determine user input intent and execute corresponding actions:}
We configure LLMs to assess user intent, directing output generation in various formats. 
Specifically, in the [Workflow] segment, we employ pseudo-code style “if” statements to prompt the LLM to identify different user intents and execute corresponding actions. In the [Output Format] segment, we utilize XML to define both the XML tags and the internal format of the content, enabling the system to process various types of content from LLMs’ output based on these tags.
For instance, the MagicDebug function must determine user intent from the input text and then take appropriate actions, such as providing text explanations or regenerating code, flowcharts, and modification nodes(see Figure~\ref{fig: systemDesign}). The system determines and executes subsequent outputs, such as re-rendering flowcharts, based on the XML tags in the output text generated by MagicDebug’s LLM module.

\textbf{How to enable LLMs to convert between code and flowcharts:}
Cocobo generates the flowchart from code and creates new code based on user modifications to the flowchart.
Cocobo uses JSON\cite{JSON} format data for front-end flowchart rendering, which is complex and lengthy due to extraneous details.
This complexity and the token length constraints of LLMs pose stability challenges.
To address these, we use Mermaid\cite{SveidqvistMermaidGeneratediagrams2014} as an intermediary flowchart description language for its simplicity and scalability.
This enables us to generate accurate and stable flowchart descriptions through prompt engineering and examples and convert them into JSON format data for frontend rendering using scripts.

\subsection{System Implementation}

Our system integrates a backend server, a web interface, and a physical robot. 
The backend is developed using Node.js, integrating OpenAI’s GPT-4 API\cite{BubeckSparksArtificialGeneral2023} to implement Cocobo’s LLM-powered functions and controlling the robot by sending WebSocket\cite{WebSocketshandbook} messages.
The frontend employs AntV X6\cite{X6JavaScriptDiagramming} for flowchart rendering and interactions. 
To evaluate the system’s usability and user perception, we used the Temi robot, which provides an SDK\cite{robotemisdk2024} that allows us to access the robot’s API and develop applications for it. 
We developed a simple robot application that launches a WebSocket server, enabling bidirectional communication between Cocobo’s backend server and the robot API. This includes, for example, controlling the robot to move to target locations, user voice input recognized by the robot, the robot’s person detection results, and so on.
The backend directly controls the robot and obtains its status, so there is no waiting time for the code to be deployed to the robot.

\section{User Study}

We conducted a user study in a lab setting to evaluate the effectiveness of Cocobo’s LLM-powered functions for novice users in customizing robot services, focusing on their perceptions of its conversational and flowchart interfaces. 
For specific task details, please refer to the Appendix.
We recruited 16 participants (8 males and 8 females), aged between 16 and 30 (mean age 22.9, SD=6.5), via social media. These participants had limited programming experience and varied familiarity with LLM-based tools.

The study comprised four phases: 1) introduction, 2) training, 3) testing, and 4) feedback. Initially, participants were briefed on the goals and structure of the test, and a sample program created by Cocobo was demonstrated. During the training phase, users watched a tutorial video and practiced using the Cocobo system under supervision. In the test phase, participants were given three tasks, each described as a scenario to program the robot accordingly. They could test the written programs at any time by deploying the code to the robot. Each task had a 25-minute time limit, but participants were allowed to continue if they needed more time. After completing the tasks, participants were asked to fill out the System Usability Scale (SUS) questionnaire and participate in a semi-structured interview.

\section{Results and Findings}
The SUS score was 72.3, indicating good usability compared to standard benchmarks \cite{AaronDeterminingwhatindividual2009}. Participants expressed confidence in using Cocobo for complex tasks, appreciating its ability to customize service robot functions. They indicated a higher likelihood of purchasing smart devices equipped with EUD tools similar to Cocobo.
All participants completed three tasks, and over half (10/16) finished within the allotted time. However, many users encountered issues, such as the LLMs failing to produce structured results as expected, generated content being too lengthy, causing delays and perceived system malfunctions, and Cocobo not meeting user expectations despite repeated input modifications. These issues suggest a need for further optimization of Cocobo’s performance.

\textbf{How do users perceive the conversational interface of Cocobo?}
Most participants (15/16) reported that the conversational interface felt natural and intelligent, enhancing the programming experience by making it seem like collaborative coding with the system.
They found the generated content aligned well with their expectations, reducing the cognitive load for those with limited programming experience. “It automatically generates it for you, which is very convenient” (P6). Additionally, it added useful interactive details. “It would add interactive details that I did not consider” (P1).
Furthermore, a subset of users (5/16) noted that clearer and more logical text inputs lead to better content generation, especially when handling complex customization requirements.

\textbf{How do users perceive the flowchart interface of Cocobo?}
The flowchart interface was generally found to be intuitive for representing code by the majority of participants (12/16), helping them quickly understand the main steps and key information without extensive reading. "The flowcharts provide specific details and direct visual representations of program steps," noted participants P3, P6, and P5. 
This feature was particularly useful for program verification and understanding, enabling users to clarify programming logic more effectively than with text descriptions alone.
However, some users (3/16) found the interface challenging without basic programming knowledge and felt that duplicated content in updates could be confusing, making it difficult to track changes. "I’m not sure where it made changes," mentioned P3.

\textbf{How do users perceive the relationship between the conversational interface and the flowchart interface?}
All users agreed that the conversational UI and flowchart UI complement each other well, utilizing different editing methods depending on the task at hand. "Some modifications are better suited for the flowchart, such as inserting a robot’s action in the process. Others are more conveniently done with text, like adjusting the overall flow," explained P8.
The Magic Debug function was utilized by only a few participants (4/16) to make precise adjustments in the flowcharts and ensure changes were applied accurately to the selected nodes, as noted by P16. However, other participants were neutral about the Magic Debug function, finding it not particularly beneficial or essential to their tasks.

\section{Limitation and Future Work}
Cocobo demonstrates the potential of integrating AI with EUD\cite{FischerAdaptiveAdaptableSystems2023} to aid everyday users in customizing smart devices and services. However, our preliminary evaluation identified several challenges that need addressing: the study only integrated basic robotic commands and did not extend to more complex IoT and network services, which may limit scalability; LLM-powered functions in Cocobo experienced issues with unstable outputs and prolonged response times due to excessively lengthy outputs; and the current design does not account for varying levels of programming skills among users, which may limit its effectiveness in practical scenarios. Moreover, this work lacks an in-depth comparison and analysis of the various representations within the Cocobo system.

For future work, we aim to enhance the performance of Cocobo's LLM-powered functions and expand the system to support additional APIs for robots and IoT devices that align with real-world application scenarios. We plan to conduct 'in-the-wild' experiments to assess the practical benefits and potential improvements of the Cocobo design concept more thoroughly. This includes exploring different representations within the system and examining EUD support for individuals across different age groups and cognitive abilities.

\bibliographystyle{IEEEtran}
\bibliography{REFERENCE}

\begin{thebibliography}{10}
\providecommand{\url}[1]{#1}
\csname url@samestyle\endcsname
\providecommand{\newblock}{\relax}
\providecommand{\bibinfo}[2]{#2}
\providecommand{\BIBentrySTDinterwordspacing}{\spaceskip=0pt\relax}
\providecommand{\BIBentryALTinterwordstretchfactor}{4}
\providecommand{\BIBentryALTinterwordspacing}{\spaceskip=\fontdimen2\font plus
\BIBentryALTinterwordstretchfactor\fontdimen3\font minus \fontdimen4\font\relax}
\providecommand{\BIBforeignlanguage}[2]{{%
\expandafter\ifx\csname l@#1\endcsname\relax
\typeout{** WARNING: IEEEtran.bst: No hyphenation pattern has been}%
\typeout{** loaded for the language `#1'. Using the pattern for}%
\typeout{** the default language instead.}%
\else
\language=\csname l@#1\endcsname
\fi
#2}}
\providecommand{\BIBdecl}{\relax}
\BIBdecl

\bibitem{LiebermanEndUserDevelopmentEmerging2006}
H.~Lieberman, F.~Patern{\`o}, M.~Klann, and V.~Wulf, ``End-user development: An emerging paradigm,'' in \emph{End User Development}, ser. Human-Computer Interaction Series, H.~Lieberman, F.~Patern{\`o}, and V.~Wulf, Eds.\hskip 1em plus 0.5em minus 0.4em\relax Dordrecht: Springer Netherlands, 2006, pp. 1--8.

\bibitem{Dragulesurveydesignspace2021}
S.~Dragule, T.~Berger, C.~Menghi, and P.~Pelliccione, ``A survey on the design space of end-user-oriented languages for specifying robotic missions,'' \emph{Software and Systems Modeling}, vol.~20, no.~4, pp. 1123--1158, 2021.

\bibitem{PaternoEnduserdevelopmentpersonalizing2019}
F.~Patern{\`o} and C.~Santoro, ``End-user development for personalizing applications, things, and robots,'' \emph{International Journal of Human-Computer Studies}, vol. 131, pp. 120--130, 2019.

\bibitem{DesoldaEmpoweringEndUsers2017}
G.~Desolda, C.~Ardito, and M.~Matera, ``Empowering end users to customize their smart environments: Model, composition paradigms, and domain-specific tools,'' \emph{ACM Transactions on Computer-Human Interaction}, vol.~24, no.~2, pp. 1--52, 2017.

\bibitem{AjaykumarSurveyEndUserRobot2022}
G.~Ajaykumar, M.~Steele, and C.-M. Huang, ``A survey on end-user robot programming,'' \emph{ACM Computing Surveys}, vol.~54, no.~8, pp. 1--36, 2022.

\bibitem{BubeckSparksArtificialGeneral2023}
S.~Bubeck, V.~Chandrasekaran, R.~Eldan, J.~Gehrke, E.~Horvitz, E.~Kamar, P.~Lee, Y.~T. Lee, Y.~Li, S.~Lundberg, H.~Nori, H.~Palangi, M.~T. Ribeiro, and Y.~Zhang, ``Sparks of artificial general intelligence: Early experiments with gpt-4,'' 2023.

\bibitem{ZhangLLaMAAdapterEfficientFinetuning2023}
R.~Zhang, J.~Han, C.~Liu, P.~Gao, A.~Zhou, X.~Hu, S.~Yan, P.~Lu, H.~Li, and Y.~Qiao, ``Llama-adapter: Efficient fine-tuning of language models with zero-init attention,'' 2023.

\bibitem{WeiEmergentAbilitiesLarge2022}
J.~Wei, Y.~Tay, R.~Bommasani, C.~Raffel, B.~Zoph, S.~Borgeaud, D.~Yogatama, M.~Bosma, D.~Zhou, D.~Metzler, E.~H. Chi, T.~Hashimoto, O.~Vinyals, P.~Liang, J.~Dean, and W.~Fedus, ``Emergent abilities of large language models,'' 2022.

\bibitem{DongSurveyIncontextLearning2023}
Q.~Dong, L.~Li, D.~Dai, C.~Zheng, Z.~Wu, B.~Chang, X.~Sun, J.~Xu, L.~Li, and Z.~Sui, ``A survey on in-context learning,'' 2023.

\bibitem{LeonardiTriggerActionProgrammingPersonalising2019}
N.~Leonardi, M.~Manca, F.~Patern{\`o}, and C.~Santoro, ``Trigger-action programming for personalising humanoid robot behaviour,'' in \emph{Proceedings of the 2019 CHI Conference on Human Factors in Computing Systems - CHI '19}.\hskip 1em plus 0.5em minus 0.4em\relax Glasgow, Scotland Uk: ACM Press, 2019, pp. 1--13.

\bibitem{ErichVisualEnvironmentReactive2017}
F.~Erich, M.~Hirokawa, and K.~Suzuki, ``A visual environment for reactive robot programming of macro-level behaviors,'' in \emph{Social Robotics}, ser. Lecture Notes in Computer Science, A.~Kheddar, E.~Yoshida, S.~S. Ge, K.~Suzuki, J.-J. Cabibihan, F.~Eyssel, and H.~He, Eds.\hskip 1em plus 0.5em minus 0.4em\relax Cham: Springer International Publishing, 2017, pp. 577--586.

\bibitem{PotChoregraphegraphicaltool2009}
E.~Pot, J.~Monceaux, R.~Gelin, and B.~Maisonnier, ``Choregraphe: a graphical tool for humanoid robot programming,'' in \emph{RO-MAN 2009 - The 18th IEEE International Symposium on Robot and Human Interactive Communication}.\hskip 1em plus 0.5em minus 0.4em\relax Toyama, Japan: IEEE, 2009, pp. 46--51.

\bibitem{PaxtonEvaluatingMethodsEndUser2018}
C.~Paxton, F.~Jonathan, A.~Hundt, B.~Mutlu, and G.~D. Hager, ``Evaluating methods for end-user creation of robot task plans,'' in \emph{2018 IEEE/RSJ International Conference on Intelligent Robots and Systems (IROS)}.\hskip 1em plus 0.5em minus 0.4em\relax Madrid: IEEE, 2018, pp. 6086--6092.

\bibitem{DattaRoboStudiovisualprogramming2012}
C.~Datta, C.~Jayawardena, I.~H. Kuo, and B.~A. MacDonald, ``Robostudio: A visual programming environment for rapid authoring and customization of complex services on a personal service robot,'' in \emph{2012 IEEE/RSJ International Conference on Intelligent Robots and Systems}.\hskip 1em plus 0.5em minus 0.4em\relax Vilamoura-Algarve, Portugal: IEEE, 2012, pp. 2352--2357.

\bibitem{PorfirioSketchingRobotPrograms2023}
D.~Porfirio, L.~Stegner, M.~Cakmak, A.~Saupp{\'e}, A.~Albarghouthi, and B.~Mutlu, ``Sketching robot programs on the fly,'' in \emph{Proceedings of the 2023 ACM/IEEE International Conference on Human-Robot Interaction}, ser. HRI '23.\hskip 1em plus 0.5em minus 0.4em\relax New York, NY, USA: Association for Computing Machinery, 2023, pp. 584--593.

\bibitem{BuchinaDesignevaluationenduser2016}
N.~Buchina, S.~Kamel, and E.~Barakova, ``Design and evaluation of an end-user friendly tool for robot programming,'' in \emph{2016 25th IEEE International Symposium on Robot and Human Interactive Communication (RO-MAN)}.\hskip 1em plus 0.5em minus 0.4em\relax New York, NY, USA: IEEE, 2016, pp. 185--191.

\bibitem{MongeRoffarelloCorrectionDefiningTriggerAction2023}
A.~Monge~Roffarello and L.~De~Russis, ``Correction to: Defining trigger-action rules via voice: A novel approach for end-user development in the iot,'' in \emph{End-User Development}, ser. Lecture Notes in Computer Science, L.~D. Spano, A.~Schmidt, C.~Santoro, and S.~Stumpf, Eds.\hskip 1em plus 0.5em minus 0.4em\relax Cham: Springer Nature Switzerland, 2023, pp. C1--C1.

\bibitem{GorostizaEnduserprogrammingsocial2011}
J.~F. Gorostiza and M.~A. Salichs, ``End-user programming of a social robot by dialog,'' \emph{Robotics and Autonomous Systems}, vol.~59, no.~12, pp. 1102--1114, 2011.

\bibitem{SaundersTeachMeShow2016}
J.~Saunders, D.~S. Syrdal, K.~L. Koay, N.~Burke, and K.~Dautenhahn, ````teach me--show me''---end-user personalization of a smart home and companion robot,'' \emph{IEEE Transactions on Human-Machine Systems}, vol.~46, no.~1, pp. 27--40, 2016.

\bibitem{CoronadoVisualProgrammingEnvironments2020}
E.~Coronado, F.~Mastrogiovanni, B.~Indurkhya, and G.~Venture, ``Visual programming environments for end-user development of intelligent and social robots, a systematic review,'' \emph{Journal of Computer Languages}, vol.~58, p. 100970, 2020.

\bibitem{GiannopoulouProgrammingItnot2021}
G.~Giannopoulou, E.-M. Borrelli, and F.~McMaster, ``"programming - it's not for normal people": A qualitative study on user-empowering interfaces for programming collaborative robots,'' in \emph{2021 30th IEEE International Conference on Robot \& Human Interactive Communication (RO-MAN)}.\hskip 1em plus 0.5em minus 0.4em\relax Vancouver, BC, Canada: IEEE, 2021, pp. 37--44.

\bibitem{OishiEndUserProgrammingRobots2017}
Y.~Oishi, T.~Kanda, M.~Kanbara, S.~Satake, and N.~Hagita, ``Toward end-user programming for robots in stores,'' in \emph{Proceedings of the Companion of the 2017 ACM/IEEE International Conference on Human-Robot Interaction}, ser. HRI '17.\hskip 1em plus 0.5em minus 0.4em\relax New York, NY, USA: Association for Computing Machinery, 2017, pp. 233--234.

\bibitem{FischerAdaptiveAdaptableSystems2023}
G.~Fischer, ``Adaptive and adaptable systems: Differentiating and integrating ai and eud,'' in \emph{End-User Development}, ser. Lecture Notes in Computer Science, L.~D. Spano, A.~Schmidt, C.~Santoro, and S.~Stumpf, Eds.\hskip 1em plus 0.5em minus 0.4em\relax Cham: Springer Nature Switzerland, 2023, pp. 3--18.

\bibitem{SarkarWhatitprogram2022a}
A.~Sarkar, A.~D. Gordon, C.~Negreanu, C.~Poelitz, S.~S. Ragavan, and B.~Zorn, ``What is it like to program with artificial intelligence?'' 2022.

\bibitem{SrinivasaRagavanGridBookNaturalLanguage2022}
S.~Srinivasa~Ragavan, Z.~Hou, Y.~Wang, A.~D. Gordon, H.~Zhang, and D.~Zhang, ``Gridbook: Natural language formulas for the spreadsheet grid,'' in \emph{27th International Conference on Intelligent User Interfaces}, ser. IUI '22.\hskip 1em plus 0.5em minus 0.4em\relax New York, NY, USA: Association for Computing Machinery, 2022, pp. 345--368.

\bibitem{Zamfirescu-PereiraWhyJohnnyCan2023}
J.~{Zamfirescu-Pereira}, R.~Y. Wong, B.~Hartmann, and Q.~Yang, ``Why johnny can't prompt: How non-ai experts try (and fail) to design llm prompts,'' in \emph{Proceedings of the 2023 CHI Conference on Human Factors in Computing Systems}, ser. CHI '23.\hskip 1em plus 0.5em minus 0.4em\relax New York, NY, USA: Association for Computing Machinery, 2023, pp. 1--21.

\bibitem{LiuWhatItWants2023}
M.~X. Liu, A.~Sarkar, C.~Negreanu, B.~Zorn, J.~Williams, N.~Toronto, and A.~D. Gordon, ````what it wants me to say'': Bridging the abstraction gap between end-user programmers and code-generating large language models,'' in \emph{Proceedings of the 2023 CHI Conference on Human Factors in Computing Systems}, ser. CHI '23.\hskip 1em plus 0.5em minus 0.4em\relax New York, NY, USA: Association for Computing Machinery, 2023, pp. 1--31.

\bibitem{JiangDiscoveringSyntaxStrategies2022}
E.~Jiang, E.~Toh, A.~Molina, K.~Olson, C.~Kayacik, A.~Donsbach, C.~J. Cai, and M.~Terry, ``Discovering the syntax and strategies of natural language programming with generative language models,'' in \emph{Proceedings of the 2022 CHI Conference on Human Factors in Computing Systems}, ser. CHI '22.\hskip 1em plus 0.5em minus 0.4em\relax New York, NY, USA: Association for Computing Machinery, 2022, pp. 1--19.

\bibitem{JiangGraphologueExploringLarge2023}
P.~Jiang, J.~Rayan, S.~P. Dow, and H.~Xia, ``Graphologue: Exploring large language model responses with interactive diagrams,'' 2023.

\bibitem{KoSixLearningBarriers2004}
A.~J. Ko, B.~A. Myers, and H.~H. Aung, ``Six learning barriers in end-user programming systems,'' in \emph{2004 IEEE Symposium on Visual Languages - Human Centric Computing}.\hskip 1em plus 0.5em minus 0.4em\relax Rome, Italy: IEEE, 2004, pp. 199--206.

\bibitem{WeiChainofThoughtPromptingElicits2023}
J.~Wei, X.~Wang, D.~Schuurmans, M.~Bosma, B.~Ichter, F.~Xia, E.~Chi, Q.~Le, and D.~Zhou, ``Chain-of-thought prompting elicits reasoning in large language models,'' 2023.

\bibitem{KojimaLargeLanguageModels2023}
T.~Kojima, S.~S. Gu, M.~Reid, Y.~Matsuo, and Y.~Iwasawa, ``Large language models are zero-shot reasoners,'' 2023.

\bibitem{VempralaChatGPTRoboticsDesign2023}
S.~Vemprala, R.~Bonatti, A.~Bucker, and A.~Kapoor, ``Chatgpt for robotics: Design principles and model abilities,'' 2023.

\bibitem{KarliAlchemistLLMAidedEndUser2024}
U.~B. Karli, J.-T. Chen, V.~N. Antony, and C.-M. Huang, ``Alchemist: Llm-aided end-user development of robot applications,'' in \emph{Proceedings of the 2024 ACM/IEEE International Conference on Human-Robot Interaction}, ser. HRI '24.\hskip 1em plus 0.5em minus 0.4em\relax New York, NY, USA: Association for Computing Machinery, 2024, pp. 361--370.

\bibitem{Ainsworthfunctionsmultiplerepresentations1999}
S.~Ainsworth, ``\BIBforeignlanguage{en}{The functions of multiple representations},'' \emph{\BIBforeignlanguage{en}{Computers \& Education}}, vol.~33, no. 2-3, pp. 131--152, Sep. 1999.

\bibitem{10.1145/3491102.3517582}
T.~Wu, M.~Terry, and C.~J. Cai, ``Ai chains: Transparent and controllable human-ai interaction by chaining large language model prompts,'' in \emph{Proceedings of the 2022 CHI Conference on Human Factors in Computing Systems}, ser. CHI '22.\hskip 1em plus 0.5em minus 0.4em\relax New York, NY, USA: Association for Computing Machinery, 2022.

\bibitem{KimMetaphorianLeveragingLarge2023}
J.~Kim, S.~Suh, L.~B. Chilton, and H.~Xia, ``Metaphorian: Leveraging large language models to support extended metaphor creation for science writing,'' in \emph{Proceedings of the 2023 ACM Designing Interactive Systems Conference}, ser. DIS '23.\hskip 1em plus 0.5em minus 0.4em\relax New York, NY, USA: Association for Computing Machinery, 2023, pp. 115--135.

\bibitem{XuInIDECodeGeneration2022}
F.~F. Xu, B.~Vasilescu, and G.~Neubig, ``In-ide code generation from natural language: Promise and challenges,'' \emph{ACM Transactions on Software Engineering and Methodology}, vol.~31, no.~2, pp. 29:1--29:47, 2022.

\bibitem{RossProgrammerAssistantConversational2023a}
S.~I. Ross, F.~Martinez, S.~Houde, M.~Muller, and J.~D. Weisz, ``The programmer's assistant: Conversational interaction with a large language model for software development,'' in \emph{Proceedings of the 28th International Conference on Intelligent User Interfaces}.\hskip 1em plus 0.5em minus 0.4em\relax Sydney NSW Australia: ACM, 2023, pp. 491--514.

\bibitem{robotemisdk2024}
\BIBentryALTinterwordspacing
``robotemi/sdk.'' [Online]. Available: \url{https://github.com/robotemi/sdk}
\BIBentrySTDinterwordspacing

\bibitem{JSON}
\BIBentryALTinterwordspacing
``{JSON}.'' [Online]. Available: \url{https://www.json.org/json-en.html}
\BIBentrySTDinterwordspacing

\bibitem{SveidqvistMermaidGeneratediagrams2014}
\BIBentryALTinterwordspacing
K.~Sveidqvist and {Contributors to Mermaid}, ``Mermaid: {Generate} diagrams from markdown-like text.'' [Online]. Available: \url{https://github.com/mermaid-js/mermaid}
\BIBentrySTDinterwordspacing

\bibitem{WebSocketshandbook}
\BIBentryALTinterwordspacing
``\BIBforeignlanguage{en}{{WebSockets} handbook}.'' [Online]. Available: \url{https://websocket.org/}
\BIBentrySTDinterwordspacing

\bibitem{X6JavaScriptDiagramming}
\BIBentryALTinterwordspacing
``\BIBforeignlanguage{en}{X6 {JavaScript} {Diagramming} {Library}}.'' [Online]. Available: \url{https://x6.antv.vision/en}
\BIBentrySTDinterwordspacing

\bibitem{AaronDeterminingwhatindividual2009}
B.~Aaron, ``Determining what individual sus scores mean: Adding an adjective rating scale,'' \emph{Journal of usability studies}, vol.~4, p.~3, 2009.

\end{thebibliography}

\appendix

\subsection{Tasks for Training}

The training session is designed to help participants become familiar with Cocobo's functions. Participants are encouraged to ask questions at any time during this stage to ensure a full understanding of the system.

Scenario 1: Design a service where the robot assists in providing guidance services. The task involves programming the robot to travel to a specified location and then return to the Reception Area.

Scenario 2: Create a robot service that greets people nearby and asks if they need navigation help. Depending on their response, the robot should provide the appropriate guidance service.

\subsection{Tasks for Formal Testing}
The formal testing consists of three tasks, each requiring the participants to conceptualize and program robot services based on provided scenarios. Participants are encouraged to debug and refine their designs until satisfaction is achieved. Each task is followed by a 5-minute break, and participants should signal the researcher upon completion of each task.

\textbf{Task 1}

Scenario: Program a robot to patrol the office after hours, reminding any remaining employees to leave. The patrol route should cover 4 specific locations and be arranged to minimize route repetition.

\textbf{Task 2}

Scenario:
With the company expanding, design a robot service to guide visitors through the exhibition area, introducing the company's projects. This includes a dual-arm robot displayed in the Exhibition Area and a mixed-reality robot testing platform in the Multimedia Studio.

\begin{enumerate}
    \item Design a tour route that includes visiting locations and an endpoint.
    \item Provide detailed introductions for each location.
    \item Plan how the robot concludes the tour.
    \item Optionally, provide different tour routes based on user responses.
    \item Optionally, include proactive service initiation by the robot.
\end{enumerate}

\textbf{Task 3}

Scenario: 
Design a service where the robot assists in guiding scheduled visitors to designated areas based on prior information provided by employees.
For example,
Miss Elaine from the administrative department has scheduled Jack to come and fix the air conditioning in the meeting room. Miss Elaine informs the robot about the task of guiding Jack to the meeting room, and when Jack arrives, the robot can guide him to the meeting room.

\begin{enumerate}
    \item Allow employees to enter visitor information, including names and destinations.
    \item Create a visitor guidance process to lead visitors to designated locations.
    \item Consider how the robot handles unknown visitors.
    \item Ensure the robot guides each visitor to the correct location.
\end{enumerate}

\end{document}